\documentclass[%
 reprint,
 amsmath,amssymb,
 aps,
]{revtex4-1}

\usepackage{graphicx}
\usepackage{dcolumn}
\usepackage{bm}
\usepackage{lineno}
\usepackage{subfigure}
\usepackage{epstopdf}

\begin{document}

\preprint{APS/123-QED}

\title{Measurement of $CP$ violation of neutral kaon system in $J/\psi$ decay\\ at the Super Tau-Charm Facility} 

\author{Jian-Yu Zhang$^{1}$}
\email{zhangjianyu@ihep.ac.cn}
\author{Cheng-Dong Fu$^{2}$}
\author{Hai-Bo Li$^{1,2}$}
\author{Liang Liu$^{3}$}
\author{Xiao-Rong Zhou$^{3}$}

\affiliation{$^{1}$University of Chinese Academy of Sciences, Beijing 100049, People's Republic of China\\
             $^{2}$Institute of High Energy Physics, Beijing 100049, People's Republic of China\\
             $^{3}$University of Science and Technology of China, Hefei 230026, People's Republic of China
}


\begin{abstract}
    In this paper, we present a preliminary study of $CP$ violation effect of $K^{0}-\bar{K}^{0}$ system in $J/\psi$ decay.
    The $CP$ violation parameters $\eta_{+-}$ and $\eta_{00}$ as well as their corresponding 
    phase $\phi_{+-}$ and $\phi_{00}$ can be determined by the difference of the time-dependent decay 
    rates between $K^{0}$ and $\bar{K}^{0}$ produced from $J/\psi \rightarrow K^{-}\pi^{+}K^{0} + c.c.$ 
    processes. We investigate the precisions of the measurements of the CP violation effect 
    at the Super Tau-Charm facility(STCF), a $e^{+}e^{-}$ collider with a peak luminosity of 
    $0.5\times10^{35}\ \rm{cm}^{-2}\rm{s}^{-1}$. 
    The parameters $\eta_{+-}$ and its phase $\phi_{+-}$ can be measured at a relative precision of 
    $1 \times 10^{-3}$, which the statistical accuracy will be several times better than that of the 
    existing PDG average values.
\end{abstract}

\maketitle


\section{\label{sec:level1}INTRODUCTION}
\begin{figure}[!h]
    \centering
	\includegraphics[width=0.8\linewidth]{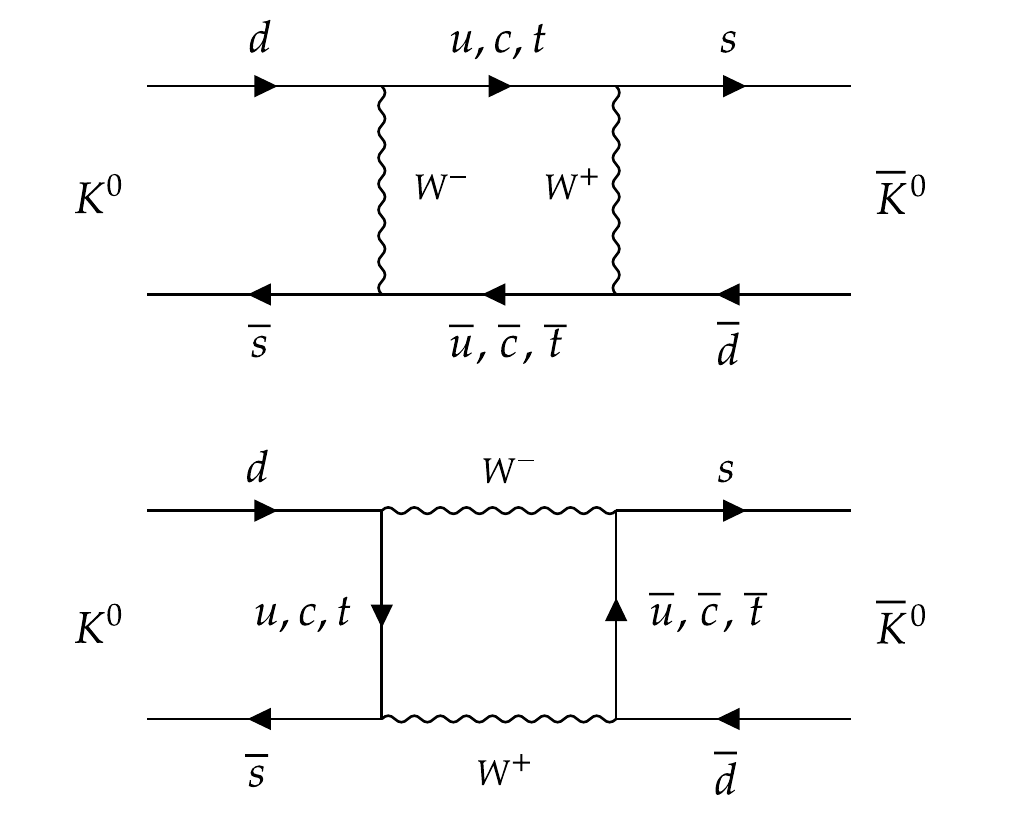}
    \caption{Feynman diagrams for the $K^{0}-\bar{K}^{0}$ mixing.}
	\label{fig:k0_k0bar_mixing}
\end{figure}

The phenomenon of mixing in neutral kaon system has been of special interest for a 
long time. Neutral kaons have definite quark components and strangeness $S$ 
when they are produced by strong interaction. The $K^{0}$ and $\bar{K}^{0}$ 
only differ by strangeness, but strangeness can be changed by flavor changing 
process(weak interaction), namely $K^{0}-\bar{K}^{0}$ mixing or oscillation. 
The Feynman diagrams of $K^{0}-\bar{K}^{0}$ mixing are shown in Fig.~\ref{fig:k0_k0bar_mixing}.
If we assume $CP$ is symmetric in weak interaction, the $CP$ eigenstates can 
be defined by $K^{0}, \bar{K}^{0}$ basis:
\begin{equation}
    \begin{aligned} 
        \left| K _ { 1  } \right\rangle & = \frac { 1  } { \sqrt { 2  }  } \left[ \left| K ^ { 0  } \right\rangle + \left| \bar { K  } ^ { 0  } \right\rangle \right ] \text {, with  } CP  = 1 \\ 
        \left| K _ { 2  } \right\rangle & = \frac { 1  } { \sqrt { 2  }  } \left[ \left| K ^ { 0  } \right\rangle - \left| \bar { K  } ^ { 0  } \right\rangle \right ] \text {, with  } CP  = - 1 . 
    \end{aligned}
\end{equation}

The decay final states with two pion or three pion have well defined $CP$ eigenvalues: 
$CP \left|\pi\pi\right\rangle = + \left|\pi\pi\right\rangle$ and 
$CP \left|\pi\pi\pi\right\rangle = - \left|\pi\pi\pi\right\rangle$. 
The phase space of $K_{1} \rightarrow \pi\pi$ is larger than 
$K_{1} \rightarrow \pi\pi\pi$, therefore the lifetime of $K_{1}$ is shorter than
$K_{2}$, so they are called $K_{S}$ and $K_{L}$. 
The $K_{S}$ and $K_{L}$ are kaon states with well defined mass 
and lifetimes that exist in nature.

However, the $CP$ symmetry is violated in weak interaction~\cite{Cabibbo:1963yz, Kobayashi:1973fv, Christenson:1964fg}. 
$K_{L}$ may also decay to $\pi\pi$ final state and $K_{S}$ 
can decay to $\pi\pi\pi$ too. It is convenient to express 
the mass states of two neutral kaons in the $\left|K_{1}\right\rangle$ 
and $\left|K_{2}\right\rangle$ basis~\cite{CPLEAR:2003ezy}:
\begin{equation}
    \begin{aligned} 
        \left| K _ { S  } \right\rangle & = \frac { 1  } { \sqrt { 1 + | \epsilon_{S} | ^ { 2  }  }  } \left( \left| K _ { 1  } \right\rangle + \epsilon_{S} \left| K _ { 2  } \right\rangle \right ) \\ 
        \left| K _ { L  } \right\rangle & = \frac { 1  } { \sqrt { 1 + | \epsilon_{L} | ^ { 2  }  }  } \left( \left| K _ { 2  } \right\rangle + \epsilon_{L} \left| K _ { 1  } \right\rangle \right ) . 
    \end{aligned}
\end{equation}
where $\epsilon_{S}$ and $\epsilon_{L}$ are complex parameters indicating possible $CP$ and $CPT$ violation. 
The parameters $\epsilon_{S}$ and $\epsilon_{L}$ can be expressed as $\epsilon_{S} = \epsilon + \delta$ and 
$\epsilon_{L} = \epsilon - \delta$ without the assumption of $CPT$ conservation;  
and $\epsilon_{S} = \epsilon_{L} = \epsilon$, if $CPT$ invariance holds. 

\begin{figure*}[!htbp]
    \centering
    \includegraphics[width=0.36\linewidth]{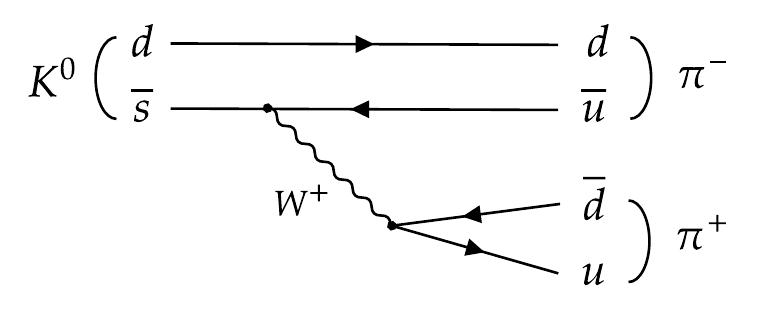}
    \hfil
    \includegraphics[width=0.36\linewidth]{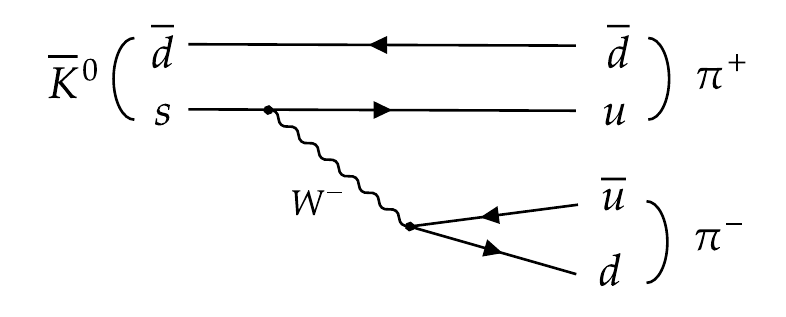}

    \includegraphics[width=0.40\linewidth]{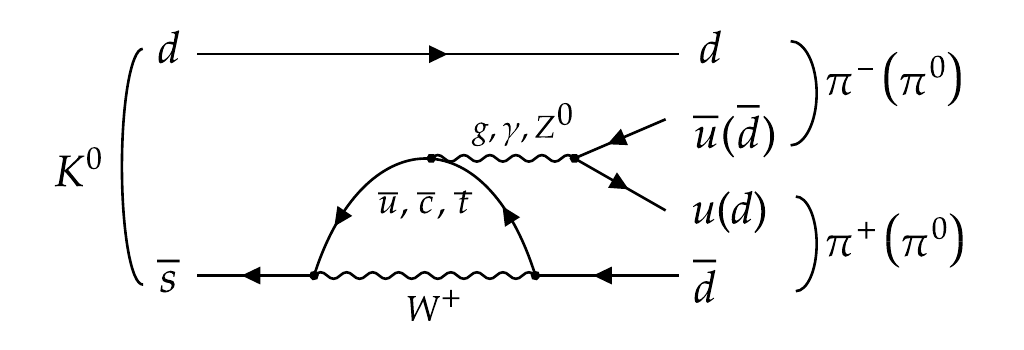}
    \hfil
    \includegraphics[width=0.40\linewidth]{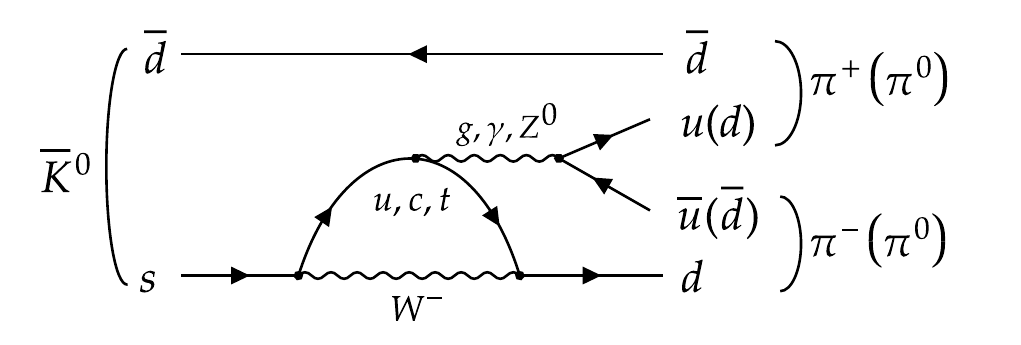}
 \caption{\label{fig:k0_penguin}%
    The Feynman diagrams at tree and penguin levels for the $K^{0}/\bar{K^{0}} \rightarrow \pi\pi$ decays.}%
\end{figure*}

There are in general three different types of $CP$ violation effects in neutral kaon decays: 
(1) indirect $CP$ violation or $CP$ violation in $K^{0}-\bar{K}^{0}$ mixing; (2) direct $CP$ violation in decay amplitude; 
(3) $CP$ violation in the interference of mixing and decay. 
We take the neutral kaon decay to $\pi\pi$ as example to explain the direct $CP$ violation. 
The $\pi\pi$ final states can be decomposed into superposition of isospin eigenstates 
$I=0$ or $I=2$. Therefore the amplitudes of a neutral kaon decay to $\pi\pi$ can be described as~\cite{CPLEAR:2003ezy}: 
\begin{equation}
    \begin{aligned} 
        A_{I} &= \left\langle \pi \pi ; I \left| \boldsymbol{H_{\mathrm{wk}}} \right| K^{0} \right\rangle , 
        \quad \bar{A}_{I} = \left\langle \pi \pi , I \left| \boldsymbol{H_{\mathrm{wk}}} \right| \bar{K}^{0} \right\rangle \\ 
        A_{I} &= \left( A_{I}+B_{I} \right ) \text{e}^{\text{i} \delta_{I}} , 
        \quad \bar{A}_{I} = \left( A_{I}^{*}-B_{I}^{*} \right) \text{e}^{\text{i} \delta_{I}} 
    \end{aligned}
\end{equation}
where $I=0,2$. The amplitudes $A_{I}$ and $B_{I}$ are $CPT$ symmetric and antisymmetric, respectively. 
The factor $\text{e}^{\text{i} \delta_{I}}$ represents the final state interaction of the pions. Direct $CP$ violation 
is arised from the phase difference between the $A_{0}$ and $A_{2}$ amplitudes. 
This phase difference is produced by the interference of tree diagrams and pengiun diagrams for $s$ quark decay, 
which is shown in Fig.~\ref{fig:k0_penguin}.

It is convenient to introduce the $CP$ violation parameters in the neutral kaon sector:
\begin{equation}
    \begin{aligned} 
        \eta _ { + -  } & = \frac { A \left( K _ { L  } \rightarrow \pi ^ { +  } \pi ^ { -  } \right )  } { A \left( K _ { S  } \rightarrow \pi ^ { +  } \pi ^ { -  } \right )  } = \left| \eta _ { + -  } \right| \text{e} ^ { i\phi _ { + -  }  } \cong \epsilon + \epsilon ^ { \prime  } \\ 
        \eta _ { 00  } & = \frac { A \left( K _ { L  } \rightarrow \pi ^ { 0  } \pi ^ { 0  } \right )  } { A \left( K _ { S  } \rightarrow \pi ^ { 0  } \pi ^ { 0  } \right )  } = \left| \eta _ { 00  } \right| \text{e} ^ { i\phi _ { 00  }  } \cong \epsilon - 2 \epsilon ^ { \prime  } , 
    \end{aligned}
\end{equation}
where $\epsilon = |\epsilon|\rm{e}^{i\phi_{\epsilon}}$ is the mixing parameter and $\epsilon^{\prime} = |\epsilon^{\prime}|\rm{e}^{i\phi_{\epsilon^{\prime}}}$ represent the direct $CP$ violation. 
At present, the PDG average value of the above $CP$ violation parameters are~\cite{ParticleDataGroup:2020ssz}: 
\begin{equation}
    \begin{aligned}
        \left| \eta _ { + -  } \right| & = ( 2.232 \pm 0.011  ) \cdot 10 ^ { - 3  } ;  & \phi _ { + -  } & = ( 43.4 \pm 0.5  ) ^ { \circ  } \\ 
        \left| \eta _ { 00  } \right| & = ( 2.220 \pm 0.011  ) \cdot 10 ^ { - 3  } ;  & \phi _ { 00  } & = ( 43.7 \pm 0.6  ) ^ { \circ  } \\ 
        | \epsilon | & = ( 2.228 \pm 0.011  ) \cdot 10 ^ { - 3  } ;  & \phi _ { \epsilon  } & = ( 43.5 \pm 0.5  ) ^ { \circ  } . \nonumber 
    \end{aligned}
\end{equation}
In which the value of $\phi_{+-}$, $\phi_{00}$ and $\phi_{\epsilon}$ are taken without assuming $CPT$ conservation.

\section{\label{sec:level1}ANALYSIS METHOD}
In this paper we first investigate the possibility of measuring the $CP$ and $CPT$ violation at a 
super $\tau-charm$ facility(STCF)~\cite{Peng:2020orp,Zhou:2021rgi}. The main part of STCF project is a symmetrical electron-positron collider 
with a peak luminosity of $0.5 \times 10^{35} \rm{cm}^{-2}\rm{s}^{-1}$ or more at center-of-mass energy $\sqrt{s} = 4.0\ \rm{GeV}$. 
The STCF will operate at the energies $\sqrt{s}$ from $2.0$ to $7.0\ \rm{GeV}$, which will provide more than 1 trillion 
$J/\psi$ events per year. Much more precise measurements of these $K^{0}-\bar{K}^{0}$ mixing and $CP$, $CPT$ violation parameters 
can be achieved at STCF. 
In experiments, there are two methods to tag the neutral kaon. The first method is $CP$ tagging method, one can 
distinguish the $K_{1}$ and $K_{2}$ by reconstruction the decay final $\pi\pi$ or $\pi\pi\pi$ states. Another 
method is flavor tagging method, the $K^{0}$ or $\bar{K}^{0}$ can be distinguished by their quark composition or 
semileptonical decay final states according to $\Delta S = \Delta Q$ rule~\cite{Feynman:1958ty}.

The method we suggest is using the charged-conjugate particles $K^{0}$ and $\bar{K}^{0}$ produced in $J/\psi$ decays. 
Initially-pure $K^{0}$ and $\bar{K}^{0}$ states are produced by decay channels: $J/\psi \rightarrow K^{0}K^{-}\pi^{+}/
\bar{K}^{0}K^{+}\pi^{-}$, with the corresponding branching ratio $(5.6 \pm 0.5) \times 10^{-3}$~\cite{ParticleDataGroup:2020ssz}. 
The strangness is conserved in the strong interaction, thus the strangeness of the neutral kaon at production 
can be tagged by the charge sign of the concomitant charged kaon. 

The $CP$ violation parameters $\eta_{+-}$ and $\eta_{00}$ can be estimated by measuring the difference of 
time-dependent decay rates between $K_{0}$ and $\bar{K}_{0}$. 
With the above definition, for the $CP$ eigenstates $f=2\pi$ or $3\pi$, 
the decay rate $R_{f}(\tau) \equiv R[K^{0}_{t=0} \rightarrow f_{t=\tau}]$ 
and $\bar{R}_{f}(\tau) \equiv R[\bar{K}^{0}_{t=0} \rightarrow f_{t=\tau}]$ 
can be written as~\cite{CPLEAR:2003ezy}:
\begin{widetext}
    \begin{equation}
        \begin{aligned}
            R_{f}(\tau) &= \frac{[1 - 2 \text{Re}(\epsilon - \delta)]}{2} \Gamma_{S}^{f} 
            \times \left[ \text{e}^{-\Gamma_{S} \tau} + \left| \eta_{f} \right|^{2} \text{e}^{-\Gamma_{L} \tau} + 2 \left| \eta_{f} \right|  
            \text{e}^{-(1/2) \left( \Gamma_{S} + \Gamma_{L} \right) \tau} \text{cos} \left( \Delta \text{m} \tau - \phi_{f} \right) \right] \\ 
            \bar{R}_{f}(\tau) &= \frac{[1 + 2 \text{Re}(\epsilon - \delta)]}{2} \Gamma_{S}^{f} 
            \times \left[ \text{e}^{-\Gamma_{S} \tau} + \left| \eta_{f} \right|^{2} \text{e}^{-\Gamma_{L} \tau} - 2 \left| \eta_{f} \right|  
            \text{e}^{-(1/2) \left( \Gamma_{S} + \Gamma_{L} \right) \tau} \text{cos} \left( \Delta \text{m} \tau - \phi_{f} \right) \right] \\ 
        \end{aligned}
    \label{eq:k0_decayrate}
    \end{equation}
\end{widetext}
where $\Gamma_{S}^{f}$ is the partial decay width of $K_{S} \rightarrow f$, 
$\Gamma_{S}$ and $\Gamma_{L}$ are the total decay width of $K_{S}$ and $K_{L}$, 
respectively. The $\Delta \text{m} = \text{m}_{L} - \text{m}_{S}$ is the mass difference between $K_{L}$ and $K_{S}$. 
The decay rate asymmetry can be formed by the combination of $R_{f}$ and $\bar{R}_{f}$: 
\begin{equation}
    \begin{aligned}
        &A_{\mathrm{CP}} ^ {f} ( \tau  ) = \frac { \bar { R  } _ {f} ( \tau  ) - R _ {f} ( \tau  )  } { \bar { R  } _ {f} ( \tau  ) + R _ {f} ( \tau  )  } \\ 
        & = 2 \rm { Re  } ( \epsilon - \delta  ) - 2 \frac { \left| \eta_{\textit{f}} \right| \mathrm { e  } ^ { ( 1 / 2  ) \left( \Gamma _ { \mathrm { S  }  } - \Gamma _ { \mathrm { L  }  } \right ) \tau  } \cos \left( \Delta m \tau - \phi _ {\textit{f}} \right )  } { 1 + \left| \eta _ {\textit{f}} \right| ^ { 2  } \mathrm { e  } ^ { \left( \Gamma _ { \mathrm { S  }  } - \Gamma _ { \mathrm { L  }  } \right ) \tau  }  } 
    \end{aligned}
    \label{eq:Acp}
\end{equation}

\section{\label{sec:level1}MONTE CARLO SIMULATION}
BESIII has collected 10 billion $J/\psi$ events which is largest so far~\cite{Asner:2008nq}, 
while the statistics are still insufficient to study $CP$ violation in neutral kaon system. 
Therfore the future STCF will be a promising facility to perform this subject. 
The STCF will accumulate about $10^{12}$ or even $10^{13}$ $J/\psi$ events per year in 
monochrome collision mode~\cite{Telnov:2020rxp}. Thus there will be more than $5.6 \times 10^{9}$ 
$K^{-}\pi^{+}K^{0} + c.c.$ events before reconstuction.
\begin{figure}[!ht]
    \centering
    \includegraphics[width=0.9\linewidth]{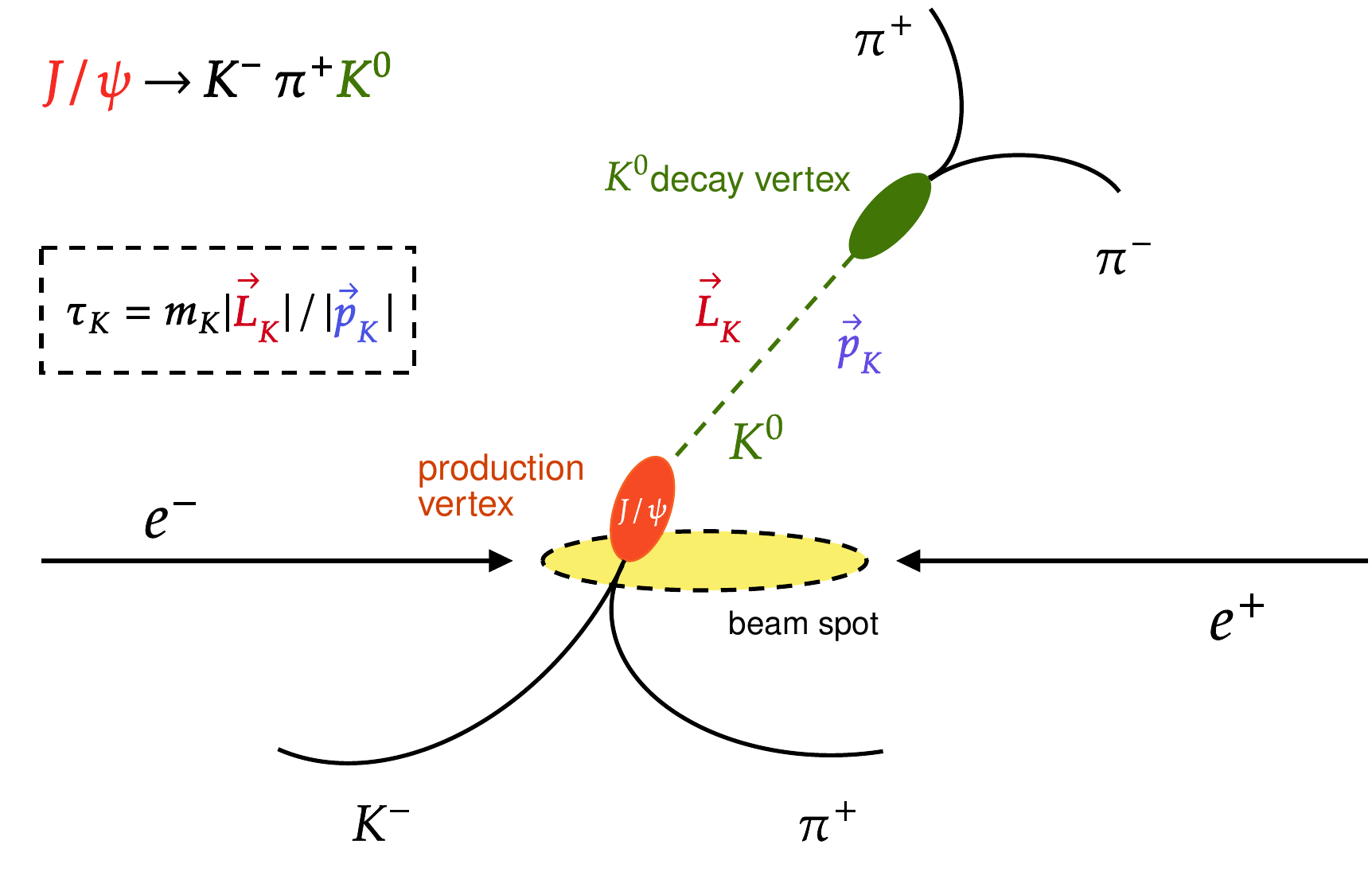}
 \caption{\label{fig:event_illustration}%
    A diagram of how to measure the decay lifetime of neutral kaon.}
\end{figure}
The method of measuring the decay lifetime of neutral kaon at STCF is 
shown in Fig.~\ref{fig:event_illustration}. 
The $J/\psi$ particle is produced after the collision of $e^{+}e^{-}$ and then rapidly decay into $K^{-}\pi^{+}K^{0}$. 
The precise measurement of the production and decay vertex of $K^{0}$
is important for the estimation of $K^{0}$ proper-time distribution.
A vertex constrained fit is performed on the $\pi^{+}\pi^{-}$ to get 
the decay vertex of $K^{0}$. The two charged pions are required to originate from a 
common vertex. And then the production vertex of $K^{0}$ is obtained by performing another vertex constrained fit on 
$K^{-}\pi^{+}$. The position uncertainty of production vertex is 
taken as $\sigma_{x} = 13.6\ \rm{\mu m}, \sigma_{y} = 1.4\ \rm{\mu m}$
and $\sigma_{z} = 50\ \rm{\mu m}$ ~\citep{Peng:2020orp,Zhou:2021rgi}, 
where the $z$ axis is along the beam direction, and the $x-y$ axes are in the plane perpendicular to the beam. 
The $\sigma_{x}$ and $\sigma_{y}$ are taken from the size of $e^{+}e^{-}$ beam in the $x-y$ 
plane, while $\sigma_{z}$ can be constrained by the vertex constrained fit of $K^{-}\pi^{+}$, that is, 
much smaller than the size of the beam in the z-direction.
The distance between $K^{0}$ production vertex and decay vertex 
is taken as the decay length $L_{K}$. 
The detector resolution for $K^{0}$ momentum $p_{K}$ is 
estimated to be $\sigma_{p} \approx 0.5\% \cdot p_{K}$.
The proper-time of $K^{0}$ can be calculated by decay length $l_{K}$ 
and momentum $p_{K}$: 
\begin{equation}
    t=\frac{m_{K^{0}l_{K}}}{p_{K}}, 
\end{equation}
where $m_{K^{0}}$ is the invariant mass of $K^{0}$. 

\begin{figure}[!htbp]
    \centering
    \includegraphics[width=0.85\linewidth]{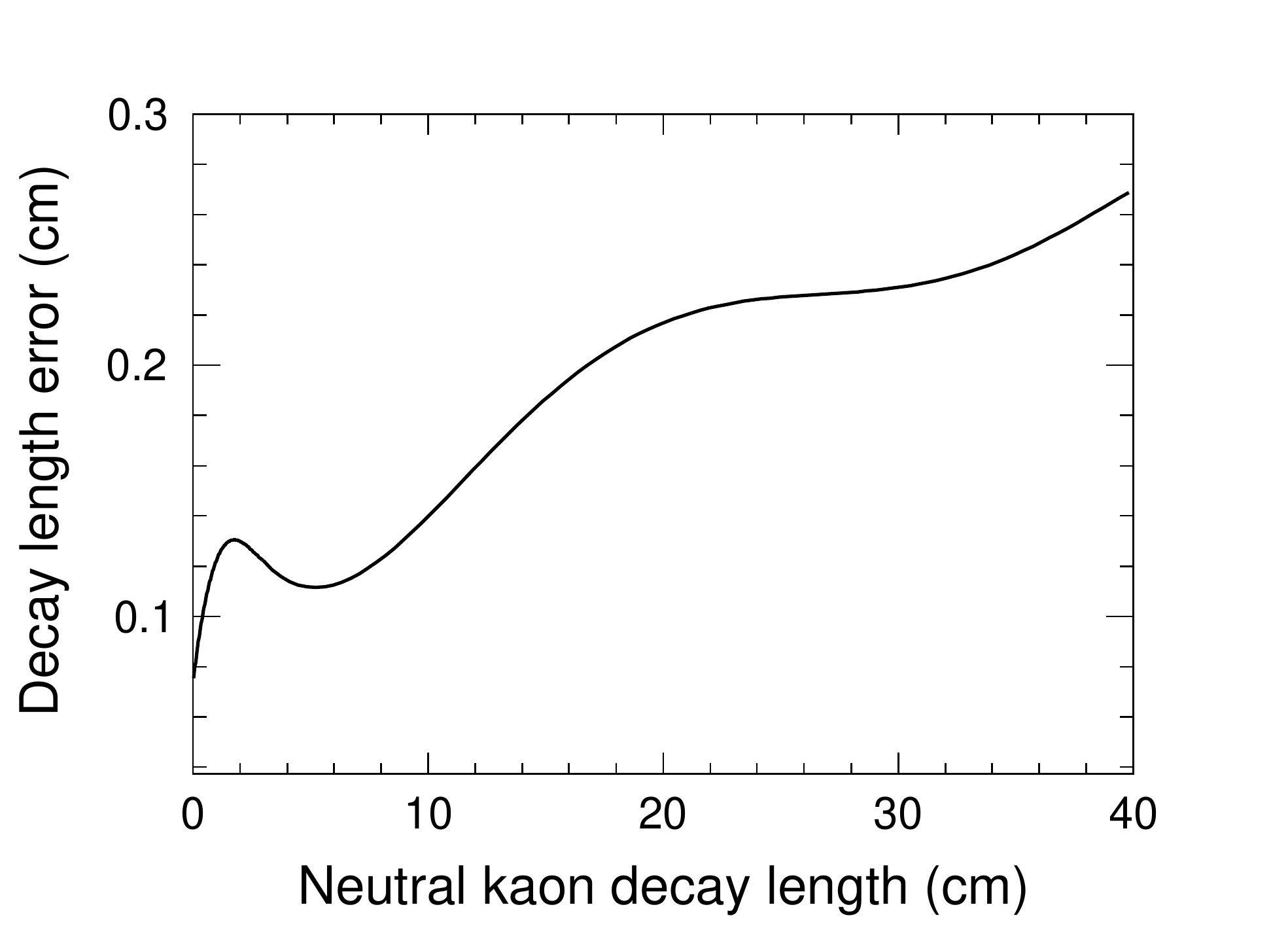}
    \hfil

    \includegraphics[width=0.85\linewidth]{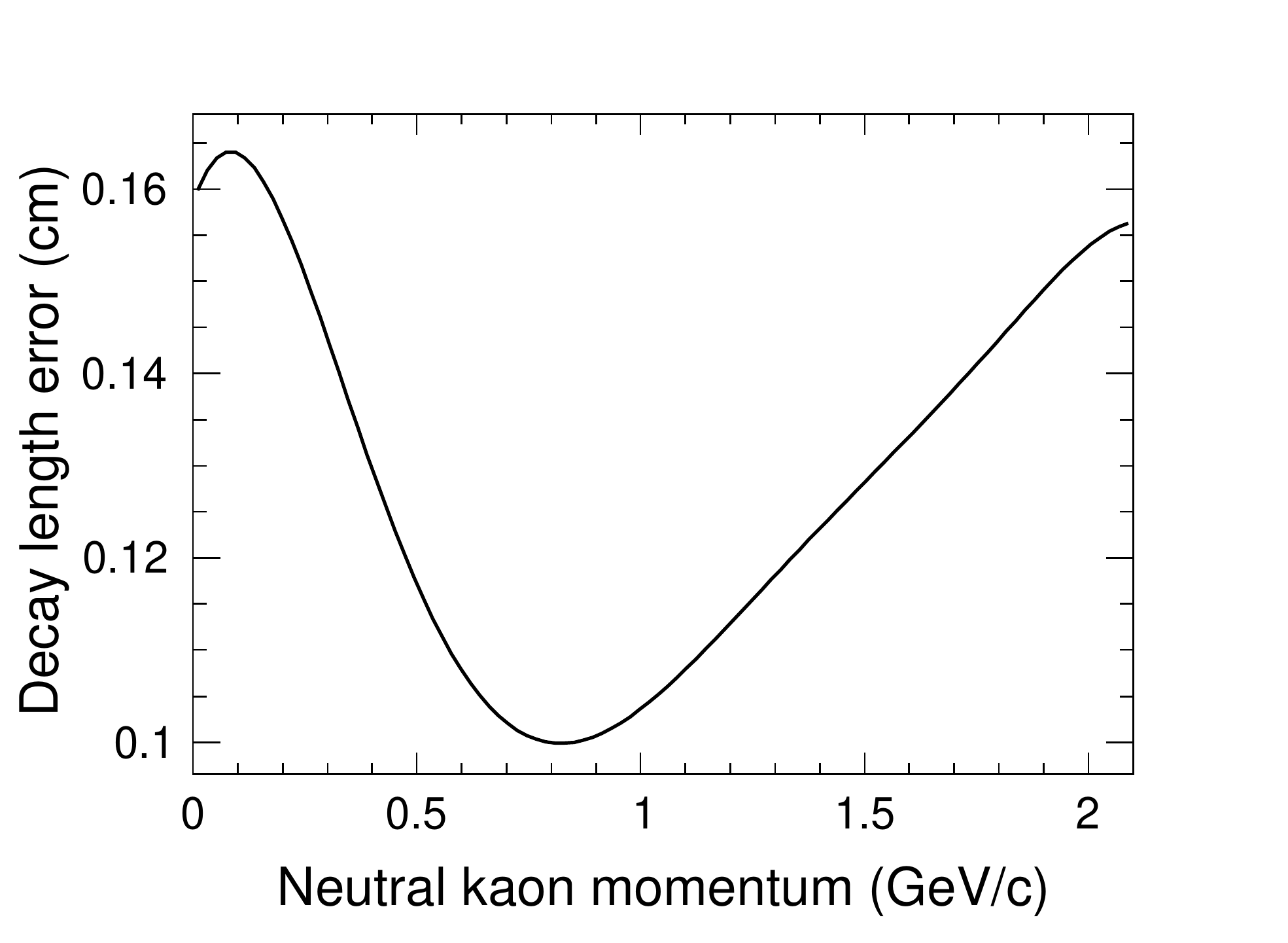}
    \caption{\label{fig:dlthE}%
    The detector resolution of $K^{0}$ decay vertex versus (a) decay length $l_{K}$ and (b) momentum $p_{K}$.}%
\end{figure}

\begin{figure*}[!htbp]
    \centering
    \includegraphics[width=0.45\linewidth]{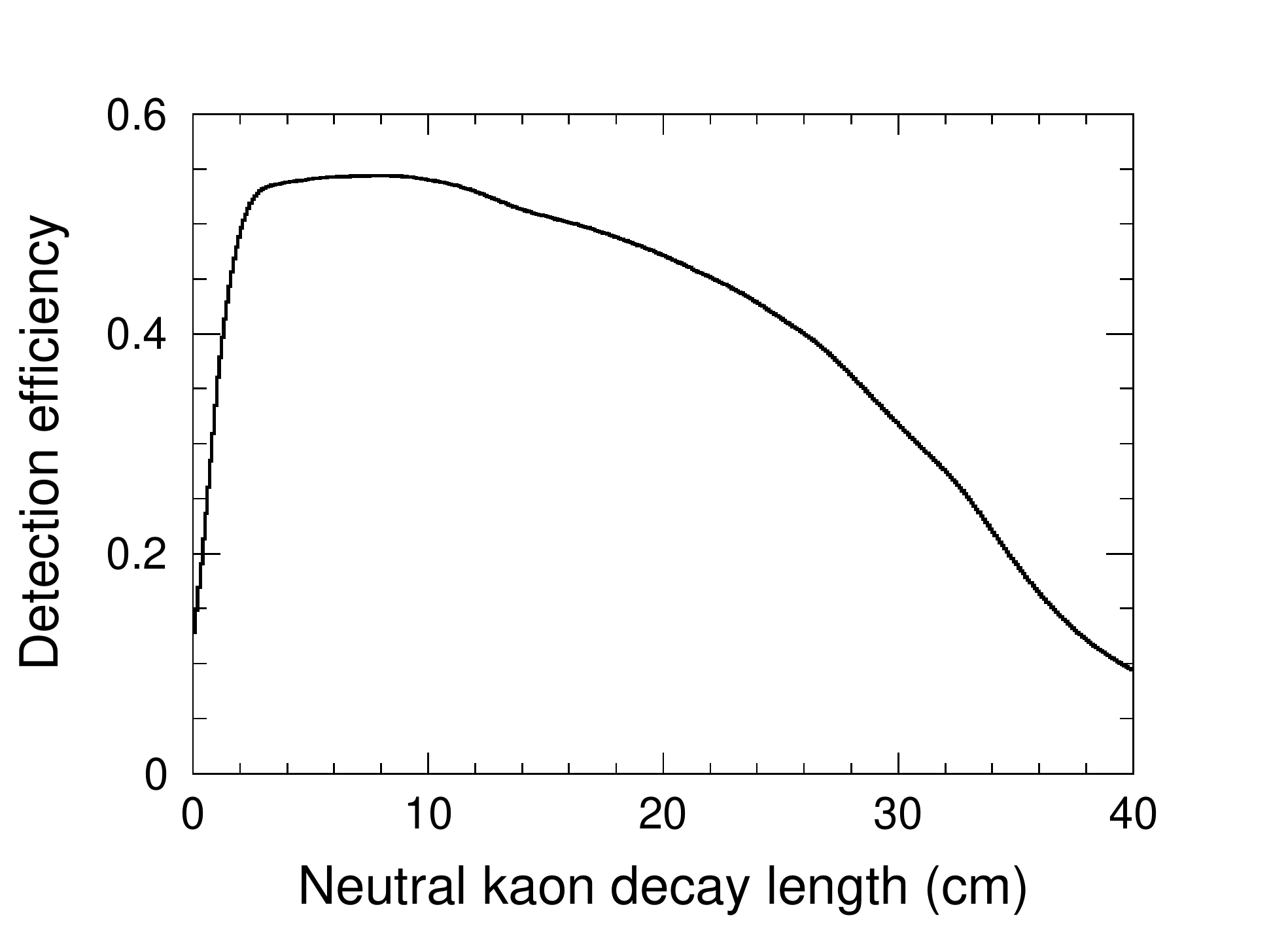}
    \hfil
    \includegraphics[width=0.45\linewidth]{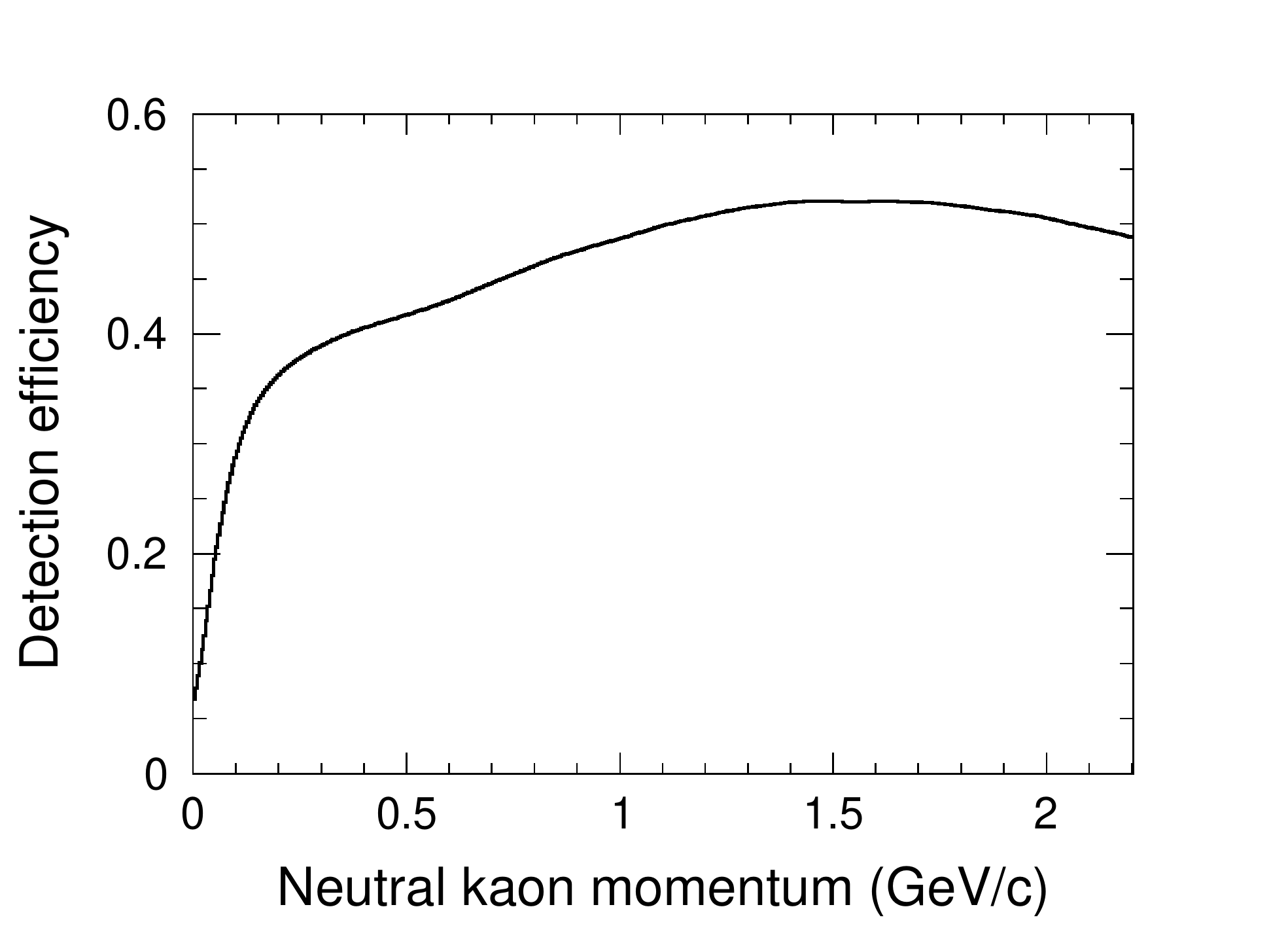}
    \caption{\label{fig:eff}%
    The detection efficiency of $K^{0}$ decay vertex versus (a) decay length $l_{K}$ and (b) momentum $p_{K}$.}%
\end{figure*}

A Monte Carlo(MC) simulation is performed on 
the basis of the total number $3.9 \times 10^{9}$ $J/\psi \rightarrow K^{-}\pi^{+}K^{0}
(K^{0} \rightarrow \pi^{+}\pi^{-}) + c.c.$ events. The time dependent decay rates of $K^{0}(\bar{K}^{0}) 
\rightarrow \pi^{+}\pi^{-}$ are simulated based on Eq.~\ref{eq:k0_decayrate} with the 
input parameters fixed to PDG average values~\cite{ParticleDataGroup:2020ssz}.
The MC simulation is generated with the considerations of the position uncertainty of $e^{+}e^{-}$ 
interaction point and $\pi^{+}\pi^{-}$ vertex reconstruction as well as the uncertainties of 
momentum reconstruction of $\pi^{+}/\pi^{-}$.
The detector resolution of $K^{0}$ decay vertex versus decay length 
as well as momentum are shown in Fig.~\ref{fig:dlthE}.
Figure ~\ref{fig:eff} shows the reconstruction efficiency versus decay length and momentum.

\begin{figure}[!ht]
    \centering
    \includegraphics[width=0.9\linewidth]{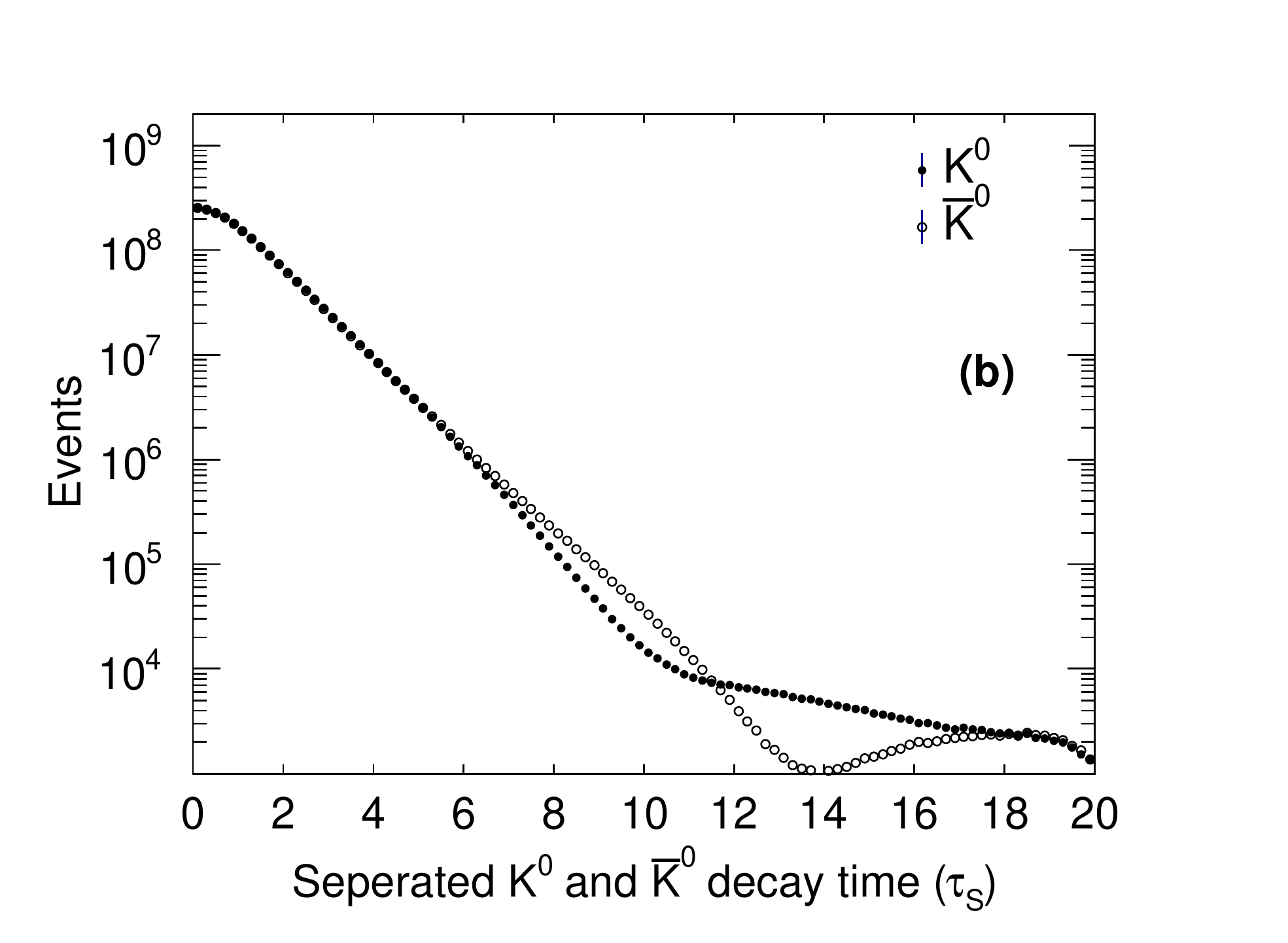}
 \caption{\label{fig:k0_decayRate}%
    The time-dependent decay rates for $K^{0}$(black point) and $\bar{K}^{0}$(open circle) separatly.}
\end{figure}

\begin{figure}[!ht]
    \centering
    \includegraphics[width=0.86\linewidth]{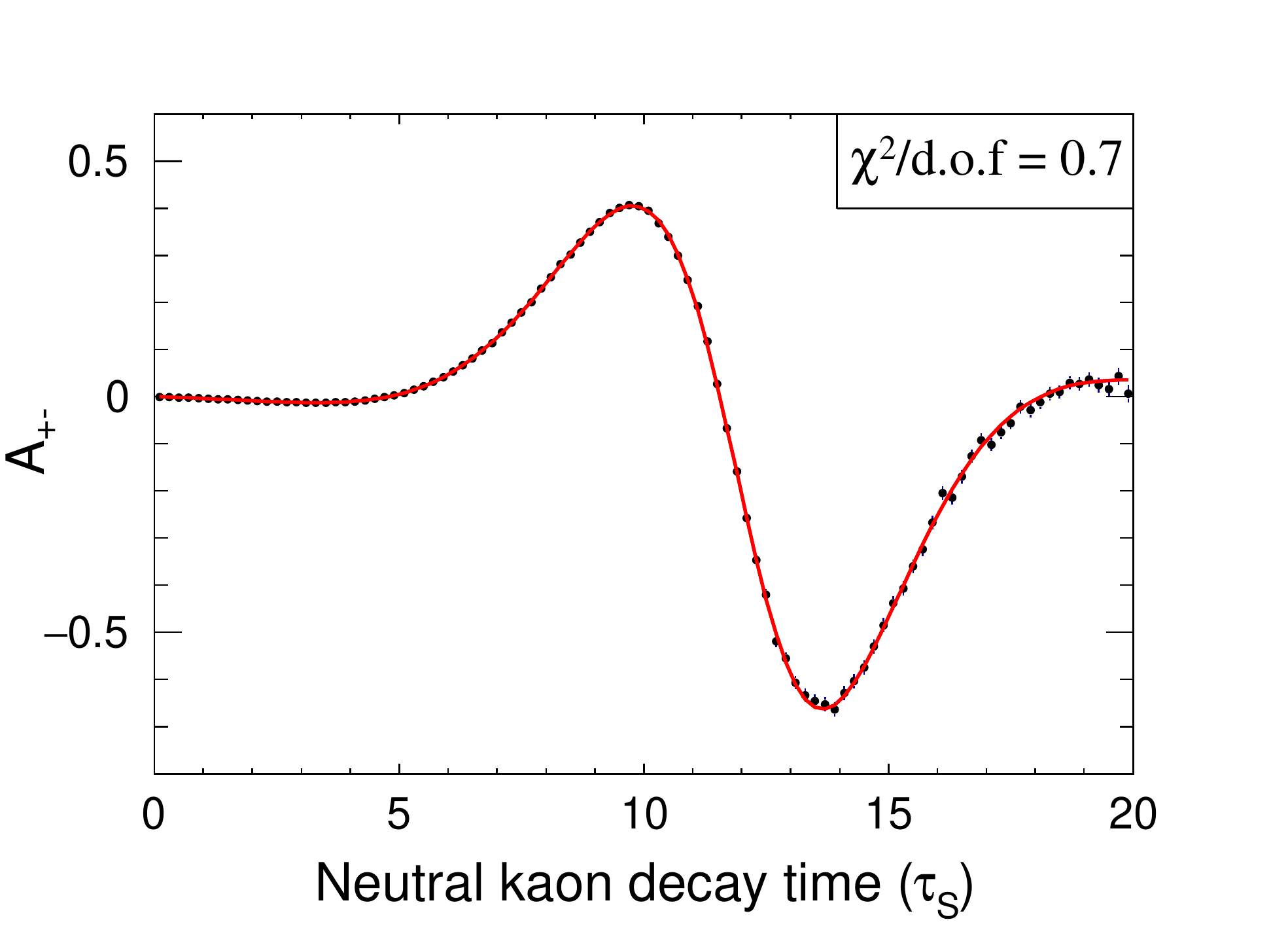}
 \caption{\label{fig:Acp}%
    The time dependent decay rate asymmetry $A_{+-}$ versus the neutral kaon decay time. The points with 
    error bar are from MC simulation and the solid line is the fitting results.}
\end{figure}

The time dependent number $N(\tau)$ for $K^{0} \rightarrow \pi^{+}\pi^{-}$ 
as well as $\bar{N}(\tau)$ for $\bar{K}^{0} \rightarrow \pi^{+}\pi^{-}$ are shown in Fig.~\ref{fig:k0_decayRate}.
Considering that this method is not sensitive to the meansurement of parameter $\rm{Re}(\epsilon-\delta)$, 
A normalization factor $k = 1+4\rm{Re}(\epsilon-\delta)$ is used to include the influence of parameter 
$\rm{Re}(\epsilon-\delta)$ on fitting~\cite{CPLEAR:1999bft}.  
Then the time dependent decay rate asymmetry $A_{+-}$ can be expressed as Eq~\ref{eq:Acp_nor}. 
We use the MINUIT method from ROOT package~\cite{Brun:1997pa} to determine the fit parameters and the fitting 
results are shown in Fig.~\ref{fig:Acp}.
\begin{equation}
    \begin{aligned}
        &A_{+-}( \tau  ) = \frac { \bar { N }( \tau  ) - k N(\tau) } {\bar { N  }( \tau   ) + k N(\tau)} \\ 
        & =  - 2 \frac { \left| \eta_{\textit{f}} \right| \mathrm { e  } ^ { ( 1 / 2  ) \left( \Gamma _ { \mathrm { S  }  } - \Gamma _ { \mathrm { L  }  } \right ) \tau  } \cos \left( \Delta m \tau - \phi _ {\textit{f}} \right )  } { 1 + \left| \eta _ {\textit{f}} \right| ^ { 2  } \mathrm { e  } ^ { \left( \Gamma _ { \mathrm { S  }  } - \Gamma _ { \mathrm { L  }  } \right ) \tau  }  } 
    \end{aligned}
    \label{eq:Acp_nor}
\end{equation}

In this fitting the parameters $\Gamma_{S}^{\pi^{+}\pi^{-}}$, $\Gamma_{S}$, $\Gamma_{L}$ 
and $\Delta\rm{m}$ are fixed to latest PDG average value~\cite{ParticleDataGroup:2020ssz}. 
The comparison between PDG average value and fitting results are shown in Table~\ref{tab:results}.
\begin{table}[htbp]
    \caption{The comparison between PDG average value and fitting results in this work.}
    \begin{tabular}{ccc}
        \hline\hline
        \makebox[0.12\linewidth]{\textbf{Par.}} & \makebox[0.5\linewidth]{$|\eta_{+-}|(10^{-3})$} & 
        \makebox[0.32\linewidth]{$\phi_{+-}(\rm{degree})$}\\
        \hline 
        \textbf{PDG} & $2.232 \pm 0.011$ & $43.4 \pm 0.5$ \\
        \hline 
        \textbf{STCF} & $2.2320 \pm 0.0025 \pm 0.0027$ & $43.510 \pm 0.051 \pm 0.059$ \\
        \hline\hline
    \end{tabular}
    \label{tab:results}
\end{table}
The first uncertainty of fitted parameters is statistical uncertainty, and the second is 
associated with the uncertainties of $e^{+}e^{-}$ interaction point and the reconstruction
of $\pi^{+}\pi^{-}$ vertex and momentum.
The fitting quality is present by $\chi^{2}/\rm{d.o.f.} = 0.7$ and the correlation 
coefficients between $|\eta_{+-}|$, $\phi_{+-}$ and normalization factor $k$ are 
shown in Table~\ref{tab:corr_coeff}.
\begin{table}
    \centering
    \caption{Correlation coefficients of the fitted parameters.}
    \begin{tabular}{c c c c}
        \hline 
        \makebox[0.2\linewidth]{\textbf{Par.}} & \makebox[0.2\linewidth]{$|\eta_{+-}|$} & 
        \makebox[0.2\linewidth]{$\phi_{+-}$} & \makebox[0.2\linewidth]{$k$} \\
        \hline
        $|\eta_{+-}|$ & 1 & -0.04 & 0.59 \\
        $\phi_{+-}$   & $-$ & 1 & 0.029 \\
        $k$ & $-$ & $-$ & 1 \\
        \hline
    \end{tabular}
    \label{tab:corr_coeff}
\end{table}

\section{\label{sec:level1}CONCLUSIONS}
To summarize, this paper present a preliminary study of $CP$ violation effects in 
neutral kaon produced at a $J/\psi$ facility. By using about $10^{12}$ $J/\psi$ events 
collected at STCF, the relative accuracy in measuring the parameters $|\eta_{+-}|$ and $\phi_{+-}$ 
is estimated to be at the level of $10^{-3}$, which will be 
several times better than the existing measurements. Measurement with such high precision needs 
a high purity data sample, accurate vertex reconstruction and well-controlled systematic uncertainties, 
which requiring STCF should have a charged track detector with excellent performance or 
a vertex detector to effectively reconstruct the neutral kaon far from the collision point. 
Especially when the proper time of $K^{0}/\bar{K}^{0}$ is between 10 and 15 times of the $K_{S}$ mean 
lifetime, there will be high sensitivity to measure $|\eta_{+-}|$ and $\phi_{+-}$. 
The corresponding decay length of neutral kaon is between 10 cm and 35 cm. Therefore, the track 
detector of STCF is required to reconstruct the vertex of $K^{0}/\bar{K}^{0}$ efficiently and accurately in this interval.
A detector with excellent position resolution and high resolution in the energy loss $dE/dx$ will help to suppress the 
background of miscombinations and particle misidentification.
The values of $\phi_{+-}$ and $\phi_{00}-\phi_{+-}$ will be used to set limits on $CPT$ violation~\cite{ParticleDataGroup:2020ssz}. 
However, the $CPT$ test by comparing the measured $\phi_{+-}$ and super weak phase $\phi_{\rm{sw}}$ need to carefully
estimate the phase difference from the contribution of the semileptonic decay modes and the $3\pi$ decay modes of neutral kaon
~\cite{Lavoura:1991nu, Nakada:1993cw, Hayakawa:1993ks, Ellis:1995xd, Ligeti:2016qpi}.
This method can also be used to study the time-dependent $CP$ asymmetries in the charmed hadron decays with 
neutral kaons in the final state at Belle(II) and LHCb~\cite{Yu:2017oky, Wang:2017gxe, LHCb:2018roe, Belle-II:2018jsg}, 
which not only can help us to study the $CP$ violation of neutral kaon system, but also 
reveal the direct $CP$ violation in the charm hadron decays after subtracting the contribution from 
neutral kaon mixing.


\begin{acknowledgments}
    We are grateful to Prof. Stephen L. Olsen for his advice and useful discussions, 
    and this work was carried out under his advice and inspiration.And the authors are grateful 
    to the STCF group for the profitable discussions. We express our gratitude to the Hefei 
    Comprehensive National Science Center for their strong support.  
    This work was supported by the National Natural Science Foundation of China (NSFC) under 
    Contracts Nos. 11935018, 11875054, 12175256; Chinese Academy of Sciences (CAS) Key Research Program 
    of Frontier Sciences under Contracts No. QYZDJ-SSW-SLH003; the international partnership program 
    of the Chinese Academy of Sciences Grant No. 211134KYSB20200057.
\end{acknowledgments}

\nocite{*}

\bibliography{apssamp}

\end{document}